\documentclass[aps,twocolumn,superscriptaddress,prb,showpacs,amsmath,amssymb]{revtex4}% Physical Review B
\usepackage{graphicx}

\begin{document}

% Use the \preprint command to place your local institutional report number on the title page in preprint mode.
% Multiple \preprint commands are allowed. \preprint{}

\title{Forming and confining of dipolar excitons by quantizing magnetic fields}
% Optional argument for running titles on pages
%\title[]{}

% repeat the \author .. \affiliation  etc. as needed \email, \thanks, \homepage, \altaffiliation all apply to the current
% author. Explanatory text should go in the []'s, actual e-mail address or url should go in the {}'s for \email and \homepage.

% \affiliation command. The \affiliation command should follow the other information

\author{K.~Kowalik-Seidl}\email[Corresponding author: ]{k.kowalik@lmu.de}
\author{X.~P.~V\"{o}gele}
\author{F.~Seilmeier}\affiliation{Center for NanoScience and Fakult\"{a}t f\"{u}r Physik, Ludwig-Maximilians-Universit\"{a}t, Geschwister-Scholl-Platz 1,
D-80539 M\"{u}nchen, Germany}

\author{D.~Schuh}\affiliation{Institut f\"{u}r Experimentelle und Angewandte Physik, Universit\"{a}t Regensburg, D-93040 Regensburg, Germany}
\author{W.~Wegscheider} \altaffiliation{Present address: Solid State Physics Laboratory, ETH Zurich, 8093 Zurich, Switzerland}
\affiliation{Institut f\"{u}r Experimentelle und Angewandte Physik, Universit\"{a}t Regensburg, D-93040 Regensburg, Germany}

\author{A.~W.~Holleitner}\affiliation{Walter Schottky Institut and Physik-Department, Technische Universit\"{a}t M\"{u}nchen, D-85748 Garching, Germany}

\author{J.~P.~Kotthaus}\affiliation{Center for NanoScience and Fakult\"{a}t f\"{u}r Physik, Ludwig-Maximilians-Universit\"{a}t, Geschwister-Scholl-Platz 1,
D-80539 M\"{u}nchen, Germany}

%Collaboration name if desired (requires use of superscript address
%option in \documentclass). \noaffiliation is required (may also be used with the \author command).
%\collaboration can be followed by \email, \homepage, \thanks as well. \collaboration{}

\date{\today}

\begin{abstract}
We show that a magnetic field perpendicular to an AlGaAs/GaAs coupled quantum well efficiently traps dipolar excitons and leads to the stabilization of the
excitonic formation and confinement in the illumination area. Hereby, the density of dipolar excitons is remarkably enhanced up to $\sim 10^{11} cm^{-2}$. By means
of Landau level spectroscopy we study the density of excess holes in the illuminated region. Depending on the excitation power and the applied electric field, the
hole density can be tuned over one order of magnitude up to $\sim 2.5$~$10^{11} cm^{-2}$ -- a value comparable with typical carrier
densities in modulation-doped structures. \\

\end{abstract}
% insert suggested keywords - APS authors don't need to do this \keywords{}

%\maketitle must follow title, authors, abstract, \pacs, and \keywords;
\maketitle

\indent The growing interest in the spectroscopy of coupled quantum wells (CQWs) is related both to the prediction of Bose-Einstein condensation of dipolar
excitons~\cite{Keldysh-ExcBecTheory, Blat-ExcBecTheory} and to the possible applications in optoelectronic devices~\cite{Krauss-photonicImages,
Zimmermann-PhotStorage, High-ExcIntegratedCircuits}. In both cases control of high densities of dipolar excitons is required. As shown both theoretically and
experimentally the dynamics of photogenerated unbound electron-hole pairs prevents the desired formation and control of a dipolar exciton
ensemble~\cite{Ivanov-thermaliz, Negoita-BlueShiftBiasedQW, Voros-diffusion, Gaertner-drift, Rapaport-NonlinearDynCQWs, Rapaport-JcondMat, Zimmermann-ExcExcTheory,
Stern-RingFormat, Voegele-PRL}. In particular excess hole densities in the illumination area of interband excitation are deduced from the formation of ring-shaped
patterns resulting from dipolar exciton transitions~\cite{Rapaport-ring, Rapaport-JcondMat, Stern-RingFormat, Hammack-ring, Paraskevov-ringBilayer}. Different
methods to spatially confine dipolar excitons within the two-dimensional (2D) plane have been applied, either by using aleatory traps created at the
interfaces~\cite{Butov-ExcTrapsTowardBEC} or by building electrostatic~\cite{Zimmermann-SLmagneticStabil, Hagn-ElFieldExcitonTransport, Hammack-ElecTraps,
Gaertner-Si0trapsSample}, stress~\cite{Negoita-stressTrap}, or magnetic~\cite{Christianen-MagnetTraps} traps. In all strategies the main goal was to create an
in-plane confining potential, which would trap dipolar excitons at its minimum. Theoretical studies of the influence of an additional magnetic field on excitonic
formation and transport suggest that a magnetic field applied perpendicular to the 2D plane will stabilize high densities of dipolar
excitons~\cite{Govorov-magnetoExcIonization, Lozovik-finiteBindirect}.

\indent We show experimentally that formation and trapping of dipolar excitons in the illumination spot is dramatically enhanced by a quantizing magnetic field.
Under focused excitation the magnetic field confines dipolar excitons at the illumination spot and thus it acts as a trap. This method allows us to achieve dipolar
exciton densities up to $\sim 10^{11} cm^{-2}$ demonstrating very high trapping efficiency, which is comparable with the most effective in-plane
traps~\cite{Hammack-ElecTraps, Gaertner-Si0trapsSample}. Additionally for higher magnetic fields we are able to determine the density of excess holes at the
excitation spot from magnetic depopulation of Landau levels, which depending on the applied electric field and excitation conditions reaches $\sim 2.5$~$10^{11}
cm^{-2}$. This also places a lower bound on the excess hole density at zero magnetic field. Presumably a high density of excess holes prevents Bose-Einstein condensation of excitons. We show how this density can be decreased using the electric field.

\begin{figure}[h]
\begin{center}
\includegraphics[width=0.5\textwidth,keepaspectratio]{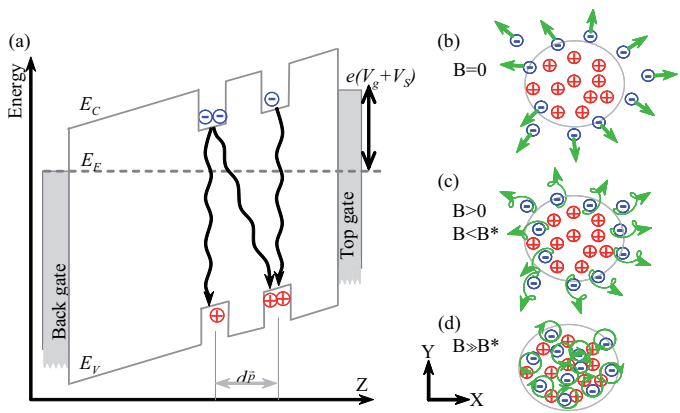}%
\end{center}
\caption{(Color online) (a) Sketch of the band gap between conduction ($E_C$) and valence ($E_V$) band edges (along the growth direction \emph{Z}). The
applied gate voltage $V_g$ and built-in voltage $V_S$ sum up to a electric field between the \emph{n}-doped back gate and the top gate. Two different optical transitions (direct and indirect) are marked with wavy arrows. (b)-(d) scheme of the in-plane (\emph{X-Y}) distribution of electrons $\ominus$ and holes $\oplus$ at the excitation spot (big circle) for different magnetic fields. Arrows mark the diffusion of electrons: (b) for $B=0$, (c) for $B<B*$, small magnetic fields, and (d) for $B \gg B*$.}\label{fig:0}
\end{figure}

\indent The studied heterostructure was grown by molecular beam epitaxy on a semi-insulating GaAs $(001)$ substrate. It consists of two $8$-nm thick GaAs coupled
quantum wells separated by a $4$-nm Al$_{0.3}$Ga$_{0.7}$As barrier and embedded between two Al$_{0.3}$Ga$_{0.7}$As barriers (Fig.~\ref{fig:0}(a)), which separte the CQWs from an \emph{n}-doped GaAs back gate and the semitransparent Schottky gate deposited on the sample surface. The resulting field-effect device allows the CQWs potential to be tuned relative to the Fermi energy pinned to the back gate. The electric field applied in the structure can be estimated from the formula: $F
=-(V_{g}+ V_{S})/d$, where $V_g$ is the voltage applied between the contacts, $V_{S}$ is the height of the metal Schottky barrier ($-0.7$~V) and $d$ is the distance
between the gates ($370$~nm). A detailed description of the structure can be found in Refs.~\cite{Gaertner-Si0trapsSample, Voegele-PRL}. The microphotoluminescence (micro-PL) spectroscopy is performed using a confocal setup~\cite{Alen-confSetup} in a He bath cryostat with a superconducting coil. A $2$~mm focal length aspheric lens is used to focus the excitation beam from a continuous diode laser ($680$~nm) and to collect the PL from the sample. Our confocal scheme allows us to investigate optical properties of the emission coming from an area of $\sim 1$~$\mu m$ in diameter~\cite{Kowalik-spinPolar}. All experiments are performed at the temperature of $4.2$~K and in magnetic fields up to $9$~T in Faraday configuration.

\indent In CQWs two types of excitonic transition can occur: a direct exciton (\emph{D}) involving an electron and a hole from the same quantum well and an
indirect exciton (\emph{I}), called also a dipolar exciton, formed by carriers from different wells (Fig.~\ref{fig:0}(a)). The dipolar transition can be easily
manipulated with the electric field (\emph{F}) and becomes energetically favorable for higher fields. Under focused excitation a spatial non-equilibrium of
photo-generated carriers forms: electrons drift out of the focus spot much faster than much heavier holes (Fig.~\ref{fig:0}(b)). Therefore the PL signal from the
illuminated area is relatively weak. The densities of electron, holes and dipolar excitons can evolve quite independently in the focus area of the illumination, a particular feature of CQWs~\cite{Rapaport-JcondMat}. In order to prevent electron escape one can use a magnetic field (\emph{B})~\cite{Govorov-magnetoExcIonization, Lozovik-finiteBindirect}. The electron drift is modified by the circular motion due to the Lorentz force resulting in a cycloid-like trajectory (Fig.~\ref{fig:0}(c)). At a threshold field $B^*$  this force dominates the electron movement, and for higher \emph{B} the electrons propagate on cyclotron orbits and stay within the excitation area (Fig.~\ref{fig:0}(d)).

\begin{figure}[t!]
\begin{center}
\includegraphics[width=0.5\textwidth,keepaspectratio]{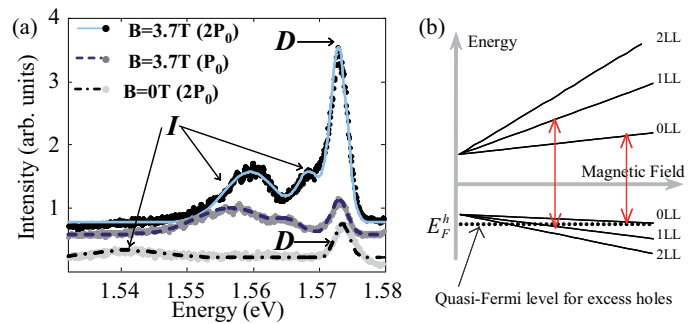}%
\end{center}
\caption{(Color online) (a) PL spectra of direct (\emph{D}) and indirect (\emph{I}) excitons for different magnetic fields and excitation powers: ($B=0$~T, $2
P_{0}$) -- lowest,  ($B=3.7$~T, $P_{0}$) -- middle, ($B=3.7$~T, $2 P_{0}$) -- upper data set ($P_{0}\approx 10 \mu W$), measured at $F=-30$~$kV/cm$. Lines --
Gaussian fits. (b) Scheme of optical selection rules between electron and hole Landau levels in a perpendicular magnetic field. Dotted horizontal line marks the
quasi-Fermi level ($E_{F}^{h}$) for holes.}\label{fig:1}
\end{figure}

\indent Figure~\ref{fig:1}(a) shows PL spectra measured at $B=0$ and $B=3.7$~T. Without a magnetic field the indirect excitons are very weak and the emission is
dominated by the direct transition (lower curve). Under applied magnetic field the intensity of the indirect transition increases by a factor of $\sim 10$ (upper
curve). Additionally, we note that the spectrum splits under magnetic fields and the optical transitions form a Landau level (LL) fan chart (Fig.~\ref{fig:1}(b)).
The two first LL transitions are clearly visible for indirect excitons (Fig.~\ref{fig:1}(a)). Comparing the spectra for two excitation powers in Fig.~\ref{fig:1}(a)
(middle and upper curves), we observe that the intensity of indirect excitons increases and their spectral position shifts toward higher energies as expected for
interacting system of dipolar excitons with increased density~\cite{Rapaport-NonlinearDynCQWs, Negoita-BlueShiftBiasedQW, Zimmermann-ExcExcTheory,
Laikhtman-BlueShift}. Also the relative intensities of the two indirect exciton LL change with laser power. The difference in the filling of consecutive LLs
suggests an increase of excess hole density with increasing excitation intensity. Already this first observation suggests the transition between hole-rich to
exciton-rich system induced by a finite $B$. In order to better understand this process we measure systematically the dependence of PL on \emph{B} under different
excitation powers and electric fields.

\begin{figure}[h]
\begin{center}
\includegraphics[width=0.5\textwidth,keepaspectratio]{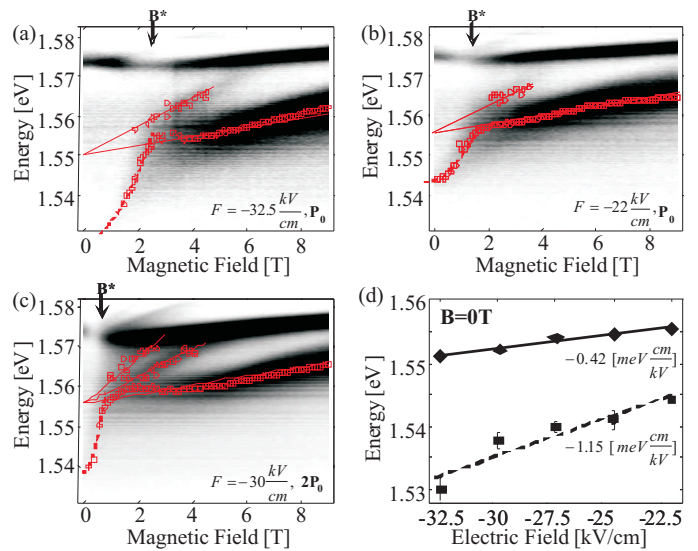}%
\end{center}
\caption{(Color online) (a)-(c) Magnetic field evolution of PL spectra for different electric fields. (a)-(b) measured at constant excitation power $P_{0}$ and (c)
-- at $2 P_{0}$. Points -- fits of energy position of LL, dashed line -- quadratic fit for low magnetic fields, solid lines -- linear fit of Landau levels for
higher fields. (d) Stark shift deduced from the extrapolation to B=0T: from the low-field evolution (squares, dashed line) and from Landau levels (diamonds, solid
line). }\label{fig:2}
\end{figure}

\indent Figures~\ref{fig:2}~(a)-(c) present the PL evolution in a perpendicular magnetic field measured for different electric fields and excitation intensities.
Two field regimes are clearly distinguishable for the indirect excitons. For low fields ($B$$<$$B^*$) the signal is relatively weak and the transition energy shifts
by several meV and approximately proportionally to $B^2$ to the blue. For higher fields, however, the emission has higher intensity, and LL splitting becomes clearly visible.

\indent The evolution of the spectra for high magnetic fields ($>$$B^*$) can be described using a simple picture of excitonic transitions in the presence of
carriers. The Landau quantization $(n+1/2)\hbar \omega _{c}$ can be fitted with cyclotron frequency $\omega _{c}=eB/m_{*}$ using a reduced excitonic mass
$m_{*}=0.05 m_{0}$, independent of $B$,$V_{g}$ and excitation intensity, and with $e$ and $m_{0}$ the electron charge and mass, respectively. The observation of
higher LL and their magnetic depopulation above a specific magnetic field $B_d$ enable us to assign a Landau level filling factor $\nu$ and hence a corresponding
effective density of excess holes $n_{h}$ via $\nu=\frac{h n_{h}}{e B_d}$, with $h$ the Planck constant. We use the first excited LL ($n$=1) to determine the dependence of $n_{h}$ on the electric field $F$ and find a systematic increase of $n_{h}$ with $F$. In Figs.~\ref{fig:2}(a)-(c) one can see the reduction and complete disappearance of this level with increasing magnetic field. We also studied higher LLs ($n$$>$1), when they were clearly resolved in the spectra, and the values of the densities obtained were the same within the experimental error. Hence, the hole density does not increase with $B>$$B^*$. We wish to point out that the excess hole density obtained from the magnetic depopulation is essentially located in the surface near the QW and at least partially excitonically bound to electrons in the lower QW. This reflects the situation where the in-plane hole distance is comparable to the spacing of the QWs which in turn is comparable to the excitonic Bohr radius.

\indent The strong blue shift of indirect PL when $B<$$B^*$ exceeds $15$~meV and is much larger than the expected diamagnetic shift of less than
$1$~meV~\cite{Bugajski-diamagShift}. However, it can be explained by the increase of the dipole-dipole interaction with increasing exciton density, in analogy to
the spectral shift with the excitation power in Fig.~\ref{fig:1}(a)~\cite{Rapaport-NonlinearDynCQWs, Negoita-BlueShiftBiasedQW, Zimmermann-ExcExcTheory,
Laikhtman-BlueShift}. At a critical magnetic field $B^*$ a maximum density is reached (see the arrows in Figs.~\ref{fig:2}(a)-(c)) and the population of the
excitons stabilizes within the excitation focus as schematically sketched in Fig.~\ref{fig:0} (c)-(d). This increase in density is also reflected in an increase in
the PL intensity. To quantify this quadratic increase in exciton density $n_{X}$ with $B$ ($B<B^*$) we subtract both the cyclotron energy and the calculated
diamagnetic shift to obtain the bare $B$-dependent and interaction caused blue shift of the indirect emission $\Delta E_{X}$ related to $\Delta n_{X}$
by~\cite{Exciton-density}:

\begin{equation}
\Delta E_{X}=\frac{e d_{\overrightarrow{p}}}{\epsilon_{b} \epsilon_{0}} \Delta n_{X} \label{eq:Xdensity}
\end{equation}

\noindent where $d_{\overrightarrow{p}}$ is the dipole length (approximately the distance between the center of the two QWs) and $\epsilon_{b}$ is the background
dielectric constant. Since the exciton density at $B=0$ is negligibly small, the magnitude of the blue shift reflects the total exciton density at $B>B^*$, which
independent of $B$ ($B>B^*$) varies from $n_{X}\sim 0.66$~$10^{11}$~cm$^{-2}$ to $n_{X} \sim 1.22$~$10^{11}$~cm$^{-2}$ for the electric field range
$-22$~to~$-32.5$~$kV/cm$. The increase of $B^*$ with the applied electric field is clearly visible in Figs.~\ref{fig:2} (a) and (b). Stronger $F$ induces larger separation of photoexcited carriers, reducing exciton binding and $n_{X}$. In order to suppress this effect higher $B$ is required to localize carriers, which allow formation of exciton~\cite{Govorov-magnetoExcIonization}. In contrast, strong excitation power increases the initial exciton density and reduces $B^*$ as seen by comparing Figs.~\ref{fig:2}(b) and (c). The $n_{X}$ enhancement in the magnetic field is a profound effect if it compares with typical densities achieved in CQWs~\cite{Voegele-PRL}.

\indent In Fig.~\ref{fig:2}(d) we compare the $B=0$ PL energy of the indirect excitons, extrapolated from the low $B$ data with that extrapolated from the Landau
level fan chart discernible at $B>B*$. For a given electric field and excitation power their energy difference reflects the blue shift induced by the $B$-dependent
change in exciton density~(\ref{eq:Xdensity}) discussed above. More striking is the change in slope in Fig.~\ref{fig:2}(d) reflecting a difference in the Stark
shift. The slope of the Stark shift obtained from the low $B$ data of $-1.15$~$\frac{meV cm}{kV}$ is in good agreement with the value $-1.26$~$\frac{meV cm}{kV}$
obtained from numerical simulations in such a heterostructure~\cite{Nextnano3} and comparable with other experiments~\cite{Gaertner-drift, Gaertner-Si0trapsSample}.
The Stark shift obtained from the extrapolation of LLs is notably reduced by about a factor of three to $-0.42$~$\frac{meV cm}{kV}$. This suggests screening
of the applied electric field, most likely caused by a magnetic-field-assisted accumulation of the dipolar indirect excitons as shown in Fig.~\ref{fig:2} ($B<B*$) and discussed above.

\begin{figure}[h]
\begin{center}
\includegraphics[width=0.5\textwidth,keepaspectratio]{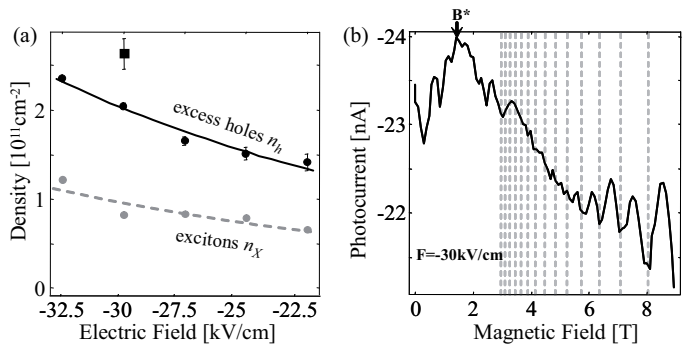}%
\end{center}
\caption{(a) Densities vs. electric field: for excess holes (black points -- series taken at excitation power $P_0$, square -- at $2 P_0$) and for the exciton
density (gray points). Solid and dashed lines are guides for the eyes. (b) Photocurrent vs. magnetic field for $F=-30$~$kV/cm$. Dashed vertical lines show the resonance of the photon energy with interband LL transition ($n=...,9,8,7$). }\label{fig:3}
\end{figure}

\indent In Fig.~\ref{fig:3} (a) we compare the excess hole $n_{h}$ and exciton $n_{X}$ densities. Note that $n_{h}$ was obtained from the depopulation of LLs ($B$$>$$B*$), whereas $n_{X}$ from the low field ($B$$<$$B*$) blue shift of PL. The changes of $n_{h}$ with the electric field can be qualitatively explained by slower escape rate of holes from the QW, which using a simple tunneling model developed in the WKB approximation~\cite{Heller-tunnelForm} and the simulated level structure in CQWs~\cite{Nextnano3} is about 3 times smaller than for electrons. The maximum $n_{h}$ exceeds $2.6$~$10^{11}$~$cm^{-2}$. We note that such high carrier density was previously observed only for intentionally doped structures. The density $n_{h}$ can exceed the density of excitons as their charge is partially compensated by the negative gate charge.

\indent Complementary information about the onset of Landau quantization can be deduced from the photocurrent measurements. Without illumination our field-effect
devices possess typical diode characteristics: current starts to flow at gate voltages above the flat-band voltage $+0.7$~V and at voltages below $-0.6$~V at
which breakdown occurs. The illumination induced photocurrent starts to flow for negative voltages and it grows with decreasing biases up to several nA at
$V_{g}=-0.5$~V ($F=-32.5$~$kV/cm$). Figure~\ref{fig:3}~(b) shows the changes of the photocurrent through the sample under applied magnetic fields. We observe a small increase of the current amplitude for $B<1.8$~T and a slow decrease for higher fields. A characteristic oscillation pattern occurs for higher \emph{B} and it is
more pronounced for bigger excitation powers. The period of the oscillations depends weakly on \emph{E}. Resonance of the LL fan chart with the laser energy
($1.82$~eV) is marked in Fig.~\ref{fig:3}~(b) by the dashed lines. When the magnetic field is increased, LLs with decreasing index cross the excitation energy,
e.g. the crossing of the seventh LL occurs at $\sim 8$~T. Because the excitation energy comes to resonance only with the empty levels and it exceeds over $20$ times
the hole quasi-Fermi energy, this effect is only seen in photocurrent measurements, but not in non-resonantly excited PL. The increase of the photocurrent at low
\emph{B} implies charge separation of photoexcited carriers, most likely occurring from within the individual QWs as reflected in the simultaneously observed
decrease in the intensity of the direct PL (see Fig.~\ref{fig:2}(a)-(c)). Note that charge separation in the respective quantum well prevents direct recombination
and thus may also increase the interwell tunneling and formation of indirect excitons. Together with the magnetic field stabilization this is likely to be
responsible for the observed intensity increase in the PL of indirect excitons.

\indent In conclusion, we showed that under focused excitation a magnetic field stabilizes the formation and confinement of indirect excitons in CQWs. The confining
ability of the magnetic field allows enhancement of the exciton densities up to $n_{X}\approx 10^{11}$~cm$^{-2}$. Our explanation is confirmed by measurements at
variable excitation intensities and electric fields. PL spectroscopy also enables us to determine the excess hole density, which can be tuned by the electric
field and the excitation power from optically non-measurable values up to $n_{h}$=$2.6$~$10^{11}$~$cm^{-2}$.

\indent The authors gratefully acknowledge helpful discussions with A. O. Govorov. This work was supported by the Center for NanoScience (CeNS), the Nanosystems
Initiative Munich (NIM), LMUexcellent, Alexander von Humboldt Stiftung, the DFG Projects KO 416/17 and HO 3324/4.

%\bibliography{BiBlio_MagExc}

\begin{thebibliography}{37}
\bibitem[{Exc()}]{Exciton-density}
\bibinfo{note}{Recently several publications were devoted to the problem of the
  exciton-exciton interaction in CQWs and correct recalculation of the blue
  shift into the exciton density~\cite{Zimmermann-ExcExcTheory,
  Schindler-ExcExcTheory, Stern-RingFormat}. The latest
  studies~\cite{Ivanov-SternComment,Laikhtman-BlueShift, Ivanov-density} show
  that for typical GaAs CQWs with exciton densities $10^{11}-10^{12}$~cm$^{-2}$
  the simple capacitor-like formula~\cite{Ivanov-thermaliz,
  BenTabou-ExcExcInteract, Rapaport-NonlinearDynCQWs} is still a good
  approximation.}
\end{thebibliography}

\end{document}